\documentclass[preprint]{aastex}
\usepackage{color}
\usepackage{graphicx}

\newcommand{\jones}{{\bf J}}
\newcommand{\stokes}{{\bf S}}
\newcommand{\gain}{{\bf G}}
\newcommand{\dterms}{{\bf D}}
\newcommand{\config}{{\bf C}}
\newcommand{\para}{{\bf P}}
\newcommand{\fara}{{\bf F}}
\newcommand{\project}{\jones_{\bf {p}}}
\newcommand{\bm}{{\bf b^{m}}_{pix}}
\newcommand{\bw}{{\bf b^{m,w}}_{pix}}

\newcommand{\bs}{{\bf b^{S}}_{pix}}
\newcommand{\bhs}{{\bf b^{S}}_{HPX}}
\newcommand{\bhats}{{\hat{\bf b}^{S}}_{HPX}}

\title{Interferometric imaging with the 32 element Murchison Wide-field Array}

\author{S.~M.~Ord\altaffilmark{1}, 
D.~A.~Mitchell\altaffilmark{1},
R. ~B.~{Wayth}\altaffilmark{1,9},  
L.~J.~Greenhill\altaffilmark{1}, 
G.~Bernardi\altaffilmark{1}, 
S.~Gleadow\altaffilmark{12}, 
R.~G.~{Edgar}\altaffilmark{14}, 
M.~A.~{Clark}\altaffilmark{1,14}, 
G.~Allen\altaffilmark{5}, 
W.~Arcus\altaffilmark{9}, 
L.~Benkevitch\altaffilmark{2}, 
J.~D. Bowman\altaffilmark{3}, 
F.~H. Briggs\altaffilmark{4}, 
J.~D. Bunton\altaffilmark{5}, 
S.~Burns\altaffilmark{6}, 
R.~J.  {Cappallo}\altaffilmark{2}, 
W.~A. {Coles}\altaffilmark{7} , 
B.~E. Corey\altaffilmark{2}, 
L. deSouza\altaffilmark{5}, 
S.~S. Doeleman\altaffilmark{2}, 
M. Derome\altaffilmark{2}, 
A. Deshpande\altaffilmark{8}, 
D.~Emrich\altaffilmark{9},
R.~Goeke\altaffilmark{9}, 
M.~R. Gopalakrishna\altaffilmark{8}, 
D.~Herne\altaffilmark{9}, 
J.~N. Hewitt\altaffilmark{10}, 
P. A. Kamini\altaffilmark{8}, 
D.~L. Kaplan\altaffilmark{15}, 
J.~C. Kasper\altaffilmark{1}, 
B.~B. Kincaid\altaffilmark{2}, 
J. Kocz\altaffilmark{4}, 
E. Kowald\altaffilmark{4}, 
E. Kratzenberg\altaffilmark{2}, 
D. Kumar\altaffilmark{8}, 
C.~J. {Lonsdale}\altaffilmark{2}, 
M.~J. Lynch\altaffilmark{9}, 
S.~R.~McWhirter\altaffilmark{2}, 
S. Madhavi\altaffilmark{8}, 
M. Matejek\altaffilmark{10}, 
M.~F. {Morales}\altaffilmark{11}, 
E. Morgan\altaffilmark{10}, 
D. Oberoi\altaffilmark{2}, 
J. Pathikulangara\altaffilmark{5}, 
T. Prabu\altaffilmark{8}, 
A.~E.~E. Rogers\altaffilmark{2}, 
A. Roshi\altaffilmark{8}, 
J.~E.~Salah\altaffilmark{2},
A.~Schinkel\altaffilmark{5}, 
N. Udaya Shankar\altaffilmark{8}, 
K.~S.~Srivani\altaffilmark{8}, 
J. Stevens\altaffilmark{5}, 
S.~J.~Tingay \altaffilmark{9}, 
A. Vaccarella\altaffilmark{4}, 
M. Waterson\altaffilmark{4,9},  
R.~L. Webster\altaffilmark{12}, 
A.~R. Whitney\altaffilmark{2}, 
A. Williams\altaffilmark{13}, 
C. Williams\altaffilmark{10}}

\altaffiltext{1}{Harvard-Smithsonian Center for Astrophysics, 60 Garden St., Cambridge MA 02138, USA (sord@cfa.harvard.edu)}
\altaffiltext{2}{MIT-Haystack Observatory, Westford MA, USA}
\altaffiltext{3}{California Institute of Technology, Pasadena, CA 91125, USA}
\altaffiltext{4}{The Australian National University, Weston Creek, ACT 2611, Australia}
\altaffiltext{5}{CSIRO Astronomy and Space Science, ATNF, POBox 76, Epping, NSW 1710, Australia}
\altaffiltext{6}{Burns Industries, 141 Canal Street, Nashua, NH 03064, USA}
\altaffiltext{7}{University of California, San Diego, La Jolla, CA 92093, USA}
\altaffiltext{8}{Raman Research Institute (RRI), Bangalore, India}
\altaffiltext{10}{MIT Kavli Institute, 70 Vassar St., Cambridge, MA 02139, USA}
\altaffiltext{11}{University of Washington,  Seattle, WA 98195, USA}
\altaffiltext{12}{University of Melbourne, School of Physics, Parkville, VIC 3010, Australia}
\altaffiltext{13}{University of Western Australia, Nedlands, WA 6009, Australia}
\altaffiltext{14}{Initiative for Innovative Computing, 29 Oxford Street, Cambridge MA 02138}

\begin{document}

\begin{abstract}

The Murchison Wide-field Array (MWA) is a  low frequency radio telescope, currently under construction, intended to search for the spectral signature of  the epoch of re--ionisation (EOR) and to probe the structure of the solar corona. Sited in Western Australia, the full MWA will comprise 8192 dipoles grouped into 512 tiles, and be capable of imaging the sky south of 40$^{\circ}$ declination, from 80\,MHz to 300\,MHz with an instantaneous field of view that is tens of degrees wide and a resolution of a few arcminutes.  A 32-station prototype of the MWA has been recently commissioned and a set of observations taken that exercise the whole acquisition and processing pipeline. We present Stokes I, Q, and U images from two $\sim$4 hour integrations of a field 20 degrees wide centered on Pictoris A. These images demonstrate the capacity and stability of a  real--time calibration and imaging technique employing the weighted addition of warped snapshots to counter extreme wide field imaging distortions.

\end{abstract}

\section{Introduction}

The Murchison Wide-field Array (MWA) is intended to be a 512 station low-frequency radio telescope, as described in detail in \cite{lcm+09}. In contrast to other connected element interferometers the MWA will not store visibilities, instead performing calibration and imaging in real--time with little or no human interaction. The MWA will be a compact radio telescope, with the longest baselines intended to be $\sim$1.5km which. This compact size is a significant advantage for treating ionospheric effects which are the dominant source of phase error. Under typical conditions the physical extent of the array is within an ionospheric isoplanatic patch, permitting image based calibration of the ionospheric phase terms  \citep{kas+08}. The baseline distribution and the time variability of the ionospheric calibration solutions requires that visibilities be averaged for no longer than 8 seconds before calibration \citep{lcm+09}, and this is the fundamental cadence of the imaging pipeline. Data volume considerations preclude the storage of 8s data cubes; the output is reduced by integrating for several minutes in the case of the EOR or restricting the field of view to a narrow range about the Sun for heliospheric science. The wide-field nature of the MWA experiment presents a significant challenge to image integration, and we  employ image-based transformations on individual snapshots as discussed in \cite{bro71}, \cite{bra84} and \cite{cp92}, in order to mitigate these effects; the authors consider this to be first time an extensive integration employing such a technique has appeared in the literature. 

A 32 tile engineering prototype (32T) intended to demonstrate  technologies required for the operation of the full array has recently been commissioned and a processing pipeline developed that is capable of calibrating the array in real-time as well as tracking, imaging and integrating in time and frequency. The outline of the paper is as follows; firstly we briefly discuss the real-time instrument calibration scheme; then we discuss the wide field imaging problem as it pertains to the MWA telescope in general and the 32 station prototype in particular; following that, image weighting and the removal of instrumental polarization are detailed; and finally  the observations are presented.

\section{Calibration}
\label{sec:cal}

The real-time calibration of the full, 512 element, MWA-512T is described in Mitchell et al. (2008)\nocite{mgw+08}  and Lonsdale et al. (2009). The
snapshot beam of the MWA-512T allows the simultaneous calibration of both the instrument gain and the ionospheric phase contribution towards many calibrator sources every 8\,seconds. The former in the form of a 2x2 complex Jones matrix for each antenna towards each calibrator, the latter a fit
for a $\lambda^2$-dependent refractive shift of each calibrator. The individual calibrator measurements are used to
constrain all-sky fits for use in the wide-field calibration of the instrument, and in the removal of strong sources
before gridding and imaging.

The snapshot sensitivity of 32T however is insufficient for making calibration measurements towards many sources, so a partial
solution has been developed. A few strong, unpolarized point sources are tracked, and  continual measurements of antenna
Jones matrices are made and the refractive effect of the ionosphere towards each source measured, and assumed to be
direction-dependent angular displacements that are constant across the field of view. We designate one calibrator -- contributing the most power -- the
\emph{primary calibrator} and use it  to set
the bulk direction-independent calibration and bandpass of each tile. Further direction-dependent gain is assumed to be
equal for each antenna and given by our default beam model. The measurement process is then
repeated for the rest of the calibrators, so that they can be subtracted if desired. We have discovered that this method is sufficient to calibrate the amplitude and phase of the tile gain at any given pointing, but have found that the amplitude of this complex gain does not transfer well to other telescope pointing positions. The phase solution is sufficiently good to permit application of one phase solution to another pointing, separated by many degrees and considerable time, but the amplitude calibration does not transfer well, being in error by several percent. 

The longest 32T baseline is $\sim 300$m and since we expect ionospheric displacements to be small compared to this relatively large 32T pixel size (Erickson, 1984)\nocite{eri84} we do not apply an ionospheric solution that is a function of position within the field of view. However, if a field contains multiple calibrators they each have an independent gain and ionospheric fit.

\label{sec:calibration}
 
\section{Wide-field Imaging}
\label{sec:wfimaging}

\subsection{The Problem}

By way of introduction, the imaging transformation performed by an interferometer can be described by the van Cittert-Zernike equation (see Clark 1973, \nocite{cla73} and Thompson, Moran and Swenson, 2001, \nocite{tms} for a detailed treatment).

\begin{eqnarray}
V(u,v,w) & = & \int \int \frac{A(\ell,m)I(\ell,m)}{n} \times \nonumber \\
&  & \hspace{-1.5cm} \exp\left(-i2\pi(\ell u+mv+(n-1)w)\right)d\ell\,dm. 
\label{eq:cz}
\end{eqnarray}

\noindent Here the $u, v$ and $w$ are the spacing (the distance between antenna pairs) measured in wavelengths, and $l,m$ and $n$ are direction cosines of the brightness distribution with respect to this coordinate frame ($n=\sqrt{1-l^{2} - m^{2}}$). The coordinate system is defined such that the $w$-axis points in the direction of the phase centre. The van Cittert-Zernike equation indicates that the complex-valued visibility function, $V(u,v,w)$, is a Fourier-like integral of the sky brightness, $I(\ell,m)$, multiplied by the primary beam response of an interferometer, $A(\ell,m)$, and $1/n$.

In {\em standard} imaging it has been conventional to make  the van Cittert-Zernike equation  independent of $n$ and $w$, and reduce it to a two-dimensional Fourier transform.   This is conventionally achieved by the the {\em small field} approximation ($n \approx 1$).  Which results in  $(n-1)w \approx 0$ and a projection from the celestial sphere onto a plane linear in $l$ and $m$, presenting an orthographic projection of the sky. However this approximation effectively neglects $w$, implying that the direction of the projection is always perpendicular to the $\ell,m$ plane, and not perpendicular to the tangent plane. This manifests as a phase error that can only be kept to a tolerable level by ensuring the small field approximation holds, thus restricting the size of the field of view. The phase error to first order is given by:

\begin{equation}
\phi_w \sim \pi w (\ell^{2} + m^{2} )
\end{equation}

\noindent The small field approximation does not hold for the MWA and therefore a mechanism must be applied to reduce this phase error; there are a number of approaches in the literature: three dimensional Fourier transforms, \citep{pc91} faceting \citep{cp92}, and w-projection \citep{cgb05}. But the MWA intends to take advantage of the fact that snapshot response of a coplanar array results in an image of the sky that is related to the true sky via a coordinate transformation (Brouw 1971;\nocite{bro71} Bracewell 1984\nocite{bra84}).

\subsubsection{Snapshot Imaging with the MWA}
\label{sec:snapshot}

As discussed in Brouw (1971)\nocite{bro71} Bracewell (1984)\nocite{bra84}, \cite{cp92} and \cite{cgb05} the co-planar assumption amounts to saying that although $w$ may not be zero, it is at all times some linear combination of $u$ and $v$, and therefore that all  the $u$ and $v$ lie in a plane given by;

\begin{equation}
w = au + bv
\end{equation}
where,  
\begin{eqnarray}
a &=& \tan{Z} \sin{\chi}\\
b &=& -\tan{Z} \cos{\chi}
\label{eqn:eta}
\end{eqnarray}
$Z$ is zenith distance, and $\chi$ is parallactic angle, as defined by.

\begin{eqnarray}
\tan{\chi} &=& \frac{\cos\phi \sin H}{\sin{\phi}\cos{\delta} - \cos{\phi}\sin{\delta}\cos H} \\
\cos{Z} &=& \sin{\phi}\sin{\delta} + 
\cos{\phi}\cos{\delta}\cos{ H},
\end{eqnarray}
where $H$ is hour angle, $\delta$ is declination and $\phi$ is latitude. We can then eliminate $w$ from  Equation\,\ref{eq:cz} to obtain:

\begin{eqnarray}
V(u,v) & = & \int \int \frac{A(\ell,m)I(\ell,m)}{n} \times \nonumber \\
&  & \hspace{-1.5cm} \exp\left(-i2\pi(\ell' u+m'v)\right)d\ell\,dm, 
\label{eq:cz_prime}
\end{eqnarray}

\noindent revealing an instantaneous Fourier relationship between sky brightness and visibility, at the expense of a position dependent coordinate distortion of the sky; that is now described by $\ell^{\prime}$ and $m^\prime$. The relationship between the measured $\ell^{\prime}$ and $m^\prime$ and the ``orthographic'' ($w=0$) $\ell$ and $m$ is given by:

\begin{eqnarray}
\ell' &=& \ell + a\left(\sqrt{1-\ell^2-m^2} - 1\right),\\
m' &=& m + b\left(\sqrt{1-\ell^2-m^2} - 1\right). \\
d\ell\,dm &=& \left(1-\frac{(al + bm)}{n} \right)d\ell'\,dm'
\label{eqn:wfe}
\end{eqnarray}

\noindent The effect of this is to warp the coordinate grid, the coordinates $\ell$ and $m$ are generally transformed to the sky coordinates of right ascension ($\alpha$) and declination ($\delta$) by the following relationships:

\begin{eqnarray}
\ell &=& \cos{\delta}\sin{\Delta\alpha} \\
m &=& \sin{\delta}\cos{\delta_0} - \cos{\delta}\sin{\delta_0}\cos{\Delta\alpha}\\
n &=& \sin{\delta}\sin{\delta_0} + \cos{\delta}\cos{\delta_0}\cos{\Delta\alpha},
\end{eqnarray}
where $(\alpha_0,\delta_0)$ is the sky position of the tracking centre and $\Delta\alpha = \alpha - \alpha_0$. These relations are then inverted to give sky coordinates:

\begin{eqnarray}
\delta &=& \arcsin\left(m\cos{\delta_0} + \sqrt{1 - \ell^{2} - m^{2}}\sin{\delta_0}\right) \\
\Delta\alpha &=& \arctan \left( \frac{\ell}{\sqrt{1 - \ell^{2} - m^{2}}\cos{\delta_0} - m\sin{\delta_0}}\right)
\label{eqn:sin}
\end{eqnarray}

\noindent If no account is taken of the difference between measured $\ell^{\prime}$, $m^\prime$ and orthographic $\ell$, $m$ then this coordinate distortion results in an incorrect labeling of the sky coordinates in the image. Conversely  this wide-field error can be negated by transforming the coordinate space of the snapshot images to account for this distortion.

It has been noted \citep{cg02} that the coordinate projection described above is a slant orthographic projection or generalised SIN-projection. This is incorporated into the FITS standard with the projection code SIN, but with extra FITS keywords to implement the slant. The SIN projection is extended with the use of the PV2\_1 and PV2\_2 FITS keywords set to $a$ and $-b$ respectively. The approach we have followed in the real-time imaging pipeline is to define a world coordinate system (WCS) for each snapshot that incorporates this projection and then  utilise  the WCSLIB coordinate transformation library to cast input pixels and their vertices into a common output frame.  The individual snapshots are accumulated in this stable frame. 

The MWA does not lie on a perfect plane, therefore a phase error will still be present due to non-coplanarity; the standard deviation from planar sampling over the tiles of the MWA-32T in the N-S direction is 0.75\,m, and in the E-W direction  0.44\,m. The topology of the array being dominated by a slope in the E-W direction of gradient\,-0.006\,mm$^{-1}$. The wavelength range of the MWA is approximately 1-3\,m. The deviation from planar sampling is indicative of the RMS deviation from zero $w$ at zenith and this corresponds to a worst case RMS phase error at the edge of a 30$^{\circ}$ FOV of  $\phi_w$ = 0.32 rad. This phase error maps to a decorrelation factor ($r$) given by the approximation:

\begin{equation}
r = (1 - \frac{1}{2}\phi_{w}^2),
\end{equation}

\noindent which is 0.95, or 5\%. For the observations presented in this paper, which have been taken at a longer wavelength than this limit, and are over a narrower FOV, the decorrelation factor is 2.5\%. 

\subsubsection{Defining the Output Frame, Resampling and Integration}

The resampling of the measured brightness distribution is the method by which the  wide-field distortions are removed in the 32T RTS imaging pipeline and the mechanism for projecting the instrument coordinates into a library frame. Due to the equal area nature of its pixels, the readily available indexing tools and the fact that it is an all sky pixelisation; we have chosen to regrid into the HEALPIX pixelisation of the sphere \citep{gor+05}. The HEALPIX scheme is commonly used by the Cosmic Microwave Background community and many tools exist to perform analysis tasks on this data format. \cite{cr07} present the family of projections to which HEALPIX belongs and also a new FITS projection (HPX) which permits imaging of HEALPIX data without any interpolation onto a regular grid.  To perform the real--time resampling we have implemented a distance-weighted interpolation scheme. Each output pixel value in the HEALPIX frame  is formed by a distance weighted average of the nearest input instrument frame pixels. Integration can be performed by the addition of the HEALPIX pixels and the application and storage of the snapshot image weights.


\subsubsection{The Weighting Scheme}
\label{sec:weighting}

The MWA tiles are phased arrays of 16 crossed--dipoles on a ground--plane, and they display significant instrumental polarization. The changing projection of the dipoles on the sky results in a polarization response that is a  strong function of source position in the telescope beam. Furthermore the tiles are not identical; different tiles display different direction-dependent complex gain due to antenna manufacturing tolerances and variations in the performance of electrical components. In combining the warped snapshots we attempt to produce a stable polarimetric measurement of the sky brightness distribution as detected by the interferometer. The changing projection of the dipoles on the sky as a function of hour angle results in different measurements of sky polarization that have varying direction-dependent weight. For example at the beginning of a track, a field rises in the east and only a fraction of the Y-polarization dipole is projected onto the sky in that direction, whereas the full X-polarization is projected. At zenith both polarizations are  presented equally to the sky. The beam response is not just a function of this projection, but  the combination of 16 phased dipoles and the ground screen. These differing contributions result in polarimetric measurements of varying fidelity as a function of hour angle, a feature that must be incorporated in the combination of snapshots by a suitable weighting scheme.

For each output image pixel we are attempting to form the best estimate of the sky polarized brightness  ($\bar{\bf b}_{pix}$) from the combination of, $n$ individual measurements (${\bf b}_{pix,j}$), each of different weight ($w_{pix,j}$). The best such combined estimate is the weighted mean:

\begin{eqnarray}
\bar{{\bf b}}_{pix} &=& \frac{\sum_{j=1}^{n} w_{pix,j} {\bf b}_{pix,j}}{\sum_{j=1}^{n}w_{pix,j}},
\label{eqn:weighted}
\end{eqnarray}
 
\noindent where the weights ($w_{pix,j}$) are generally the reciprocal of the pixel variance. The inverse-variance of the MWA image pixels is best represented by the direction dependent polarimetric tile response in the direction of that pixel (the power response). We are weighting therefore by the power signal to noise ratio, under the assumption that the the thermal noise is direction independent. Following the formalism developed in the series of papers: \cite{hbs96}; \cite{shb96}; \cite{hb96}; \cite{ham00} and \cite{ham06}, the brightness of an image pixel in the measured, or instrument frame is:

\begin{equation}
\bm =  (\jones_{pix} \otimes \jones_{pix}^{*})\, \stokes\, \bs,
\end{equation}

\noindent where ${\bf b}$ are the coherency 4-vectors, the superscript ${\bf m}$ indicates ``measured'' and should be taken to imply an instrument polarization basis ($p,q$) and ${\bf s}$ represents a  Stokes ($I, Q, U, V$) basis. The matrix $\stokes$ transforms polarization from the instrument basis (orthogonal linear feeds) to the Stokes basis on the sky and is

\begin{equation}
 {\bf S} = \frac{1}{2} \left( \begin{array}{cccc}
  1 & 1 & 0 & 0 \\
  0 & 0 & 1 & i \\
  0 & 0 & 1 & -i \\
  1 & -1 & 0 & 0 \\
  \end{array} 
  \right).
 \end{equation}

\noindent  The MWA-32T is an interferometer comprising 32 different elements but, as is described in \S \ref{sec:calibration}, each visibility is calibrated by an equalization process that allows us to approximate the instrument polarimetric response by an {\em instrument Jones matrix}. The 2x2   instrument Jones matrix    $\jones$, determines how the generally non-orthogonal projected receptors ({\em p,q}) respond to the incident orthogonal sky polarization (X, Y).  As the projection of the tile receptors changes as a function of position the Jones matrix is different for every pixel. 

It is this ``instrument Jones matrix'', that is used to form the variance weight matrix for each input pixel. As we are not accounting for the different primary beam shape, we are not forming the best estimate of the array response in a given pixel direction, it is this simplification  that makes the problem tractable in the 32T domain. But it should be  noted that the weighting scheme will evolve as the MWA develops. As more tiles are added, significantly over conystraining the parameters of the array, a direction dependent calibration scheme as described in \cite{mgw+08}, with a more accurate weighting scheme incorporating the differences between tiles, will be applied in the visibility domain.

\subsubsection{Building and Applying the Weights}
 
The 2x2 Jones matrix of a radio telescope is generally decomposed into: 

\begin{equation}
\jones = \gain \dterms \config \para \fara,
\end{equation}
\noindent where $\gain$ is the electronic gain, $\dterms$ represent the feed errors, $\config$ is the configuration of the feed response, $\para$ is parallactic rotation and $\fara$ is Faraday rotation. We have not attempted to correct for Faraday rotation in the experiment, the ramifications of which are explored in \S \ref{sec:stokes}. In our application the dipoles do not rotate relative to the source, but the projection of the dipoles changes on the sky. 

\begin{equation}
\jones = \gain \dterms \project .
\end{equation}

\noindent Where $\project$ is the dipole projection matrix given by
 \begin{equation}
 \project = \left(  \begin{array}{cc}
 \cos{L}\,\cos{D} + \sin{L}\,\sin{D}\,\cos{H} & -\sin{L}\,\sin{H} \\
 \sin{D}\,\sin{H} & \cos{H}\\ 
 \end{array} \right)
 \end{equation}
 
\noindent $L$, $D$ and $H$ are latitude, declination and hour angle respectively. In the calibration stage of the imaging pipeline \citep{mgw+08} the Jones matrices of the individual antenna elements are solved for, which essentially allows $\gain$ and $\dterms$ to be obtained, but they are incorporated in the fitted Jones matrix and not decomposed. 

The 32T calibration scheme attempts to solve for  the elements of the diagonal $\gain$ matrix and the rest of the matrices are assumed to be the same for all tiles. Thus individual beams can be scaled and shifted (the elements are complex) but not changed in shape. This is required by the limited sensitivity of the 32T array. 

The weight matrix must be constructed for every input sky pixel and in the 32T RTS is pre-computed at the beginning of each $\sim$3 minute integration. For each pixel the image coordinates ($\ell^{\prime}$ and $m^\prime$) are converted into a topocentric hour angle and declination. In this process account has to be taken of the initial orthographic projection, and the required spherical coordinates. The model of the beam is then formed as the sum of 16 dipoles with a phase delay commensurate with the required tile pointing centre. The complex gain of the tile in the desired pixel direction is then calculated and the Jones matrix for that pixel ($ \jones_{pix}$) formed.

The next step is the formation of the image weights, ideally these  would variance weight the image so as to provide close to a maximum likelihood estimation of the sky. We do not have the true power response of the instrument, but we have the fitted model response $\jones_{pix}$. We therefore calculate the power beam response from the fitted voltage response, and this is done by forming the outer product ($\jones_{pix} \otimes \jones_{pix}^{*}$). We then form the Hermitian transpose of this matrix. And each input pixel polarimetric vector is formed from the measured instrumental polarization ($\bm$) multiplied by the weight matrix:

\begin{eqnarray}
\bw &=& (\jones_{pix} \otimes \jones_{pix}^{*})^{\dagger} (\jones_{pix} \otimes \jones_{pix}^{*})\, \stokes\, \bs.
\label{eqn:weighted2}
\end{eqnarray}

\noindent This is performed for every pixel in each snapshot. Then each input pixel is resampled into the output HEALPIX frame.

\section{Stokes Conversion}
 \label{sec:stokes}
Accounting for the changing response of the tiles to correctly weight the instrumental polarization maps permits the production of a best estimate of the Stokes parameters when the weighted images are correctly normalized. However the observations that are presented here were not taken in a manner that maximizes polarization fidelity. For the purposes of clarity we will emphasize the following points:
\begin{itemize}
\item{We are assuming a direction dependent response for the instrument for which a complex gain has been determined, but it is the same for each tile. This is not the case and each tile will make a different contribution to the array response. This mimics a leakage between Stokes parameters at a level commensurate with the departure from the {\em mean} beam shape.}
\item{The bandwidth of each channel of the presented images is 1.92~MHz wide, and will inflict considerable bandwidth depolarization on celestial sources: from \cite{gw66} the depolarization factor is given by $\sin{\Delta\phi}/\Delta\phi$, where $\Delta\phi$ is given by:
\begin{equation}
\Delta\phi \sim - \mathrm{RM} \lambda_{0}^{2} \frac{2\Delta\nu}{\nu},
\end{equation}
at a central wavelength of 2m, this factor is 1\% at an RM of 1\,radm$^{-2}$, but 100 \% at an RM of $\sim$10\,radm$^{-2}$. But we are also not correcting for any Faraday rotation between these channels, which will effectively ($\sim$80\%) depolarise a signal with an RM as low as 1\,radm$^{-2}$. } 
\item{Another depolarization factor results from not accounting for any variations in Faraday rotation within and between the sub-integrations. The integrations run for several hours and are made up of 3~minute sub-integrations, \cite{epfk01} present compelling observation at 327~MHz with the VLA that demonstrate Faraday rotation of 10's of degrees throughout several hours of daytime observations. Our observations are at lower frequencies so we would expect even more rotation (a factor of $\sim$ 4 more) rotating flux between Stokes Q and Stokes U throughout the observation. The observations presented here were taken at night-time which should mitigate this effect, however observations of the pulsar J0218+4232 at the Westerbork Synthesis Radio Telescope found typical night-time RM variations of 0.2-0.4 radm$^{-2}$, about 70 degrees of variation at our frequencies (G. Bernardi unpublished). }
\end{itemize}

\noindent Even a benevolent ionosphere with RM deviations considerably less than 1 radm$^{-2}$ would be still be sufficient to significantly depolarise any polarized source, or diffuse background during the long integrations presented here. Given the combination of these various de-polarization effects any residual polarization in the maps presented here are most likely due to instrumental polarization leaking between the Stokes parameters due to imprecise beam models, and polarized sidelobes from bright, intrinsically unpolarized, diffuse emission from outside the field of view.

\subsection{Reconstructing the Weights and Converting to Stokes Parameters}

Each output sky pixel is the result of the combination of many pixels with different weights, applied at different times. The 32T prototype solution to this problem is to store the parameters that are used to analytically calculate the weights.  This information is then used offline to reconstruct the direction dependent instrument Jones matrix for each constituent weight, at the time and sky-position that each weight was applied.  

\subsubsection{The Stokes Conversion as Part of Image Normalisation}
\label{sec:norm}
The weight normalization factor would normally be the sum of the variance weights as per Equation \ref{eqn:weighted} however in order to simultaneously normalize the images and convert to Stokes parameters from instrument polarization the following product is formed:

\begin{eqnarray}
 {\bf N}_{HPX} &=& \sum_{j=1}^{n} (\jones_{j,pix} \otimes \jones_{j,pix}^{*})^{\dagger} (\jones_{j,pix} \otimes \jones_{j,pix}^{*})
\label{eqn:norm}
\end{eqnarray}

\noindent It is the formation of this product that permits the transformation to Stokes parameters. Essentially the images are being stored in weighted instrument polarization, but the information required to convert the instrumental polarization to the X-Y basis is also being stored. This is reasonable as the weights are formed from the beam model, which has a limited number of parameters, if the weight matrix had to be stored then 4 times more weight information would have to be stored than actual measurements.

The weight and transformation product is evaluated, summed and inverted for every pixel. We then multiply the weighted pixels by this normalization matrix to produce an estimate of the true sky brightness $\bhats$ in the following manner:

\begin{equation}
\bhats =  \sum_{j=1}^{n}\left[ (\jones_{j,pix} \otimes \jones_{j,pix}^{*})^{\dagger} (\jones_{j,pix} \otimes \jones_{j,pix}^{*})\right]^{-1} . \\ \sum_{j=1}^{n}\left[(\jones_{j,pix} \otimes \jones_{j,pix}^{*})^{\dagger} (\jones_{j,true} \otimes \jones_{j,true}^{*})\right] \, \stokes\, \bhs.
\end{equation}

\noindent As outlined above, in practice the summations occur during the image integration step and the weight integration step, so dropping the summation and the $j$ subscript for clarity we have:

\begin{equation}
\bhats = \left[(\jones_{pix} \otimes \jones_{pix}^{*})^{\dagger} (\jones_{pix} \otimes \jones_{pix}^{*})\right]^{-1}(\jones_{pix} \otimes \jones_{pix}^{*})^{\dagger} (\jones_{true} \otimes \jones_{true}^{*})\, \stokes\, \bhs,
\end{equation}

\subsubsection{On the Equality of $\jones_{true}$ and $\jones_{pix}$}
\label{sec:jpx}

The operation described above effectively cancels the instrument response, but only to the extent that the $\jones_{pix}$ represents $\jones_{true}$ as discussed in \S \ref{sec:weighting}.    It is the level to which this does not hold that constrains the polarization fidelity of this experiment. Under this assumption the application of $\stokes^{-1}$ will convert the normalized sky coherency vector from the X-Y basis to the Stokes basis. 

The tile model is a simplification of the true tile response, although informed by EM simulations there is surely an appreciable difference in the polarized response of the true tile, and that predicted by the current tile model. This is another aspect of 32-T processing that will be less of a factor for the full array, which will attempt to fit more parameters of the tile model than has been employed here. Over and above model errors and the tile to tile variations there are other possible differences between  $\jones_{true}$ and $\jones_{pix}$, an important one being Faraday rotation. As discussed in both \S \ref{sec:weighting} and \S \ref{sec:stokes} the Faraday rotation is included in the Jones decomposition of the tile response, however it is not included in $\jones_{pix}$. As we know there is an unknown Faraday component to $\jones_{true}$, flux will be redistributed between Q and U as a function of time, which will effectively depolarise the images over long integrations.  This effect can be corrected by including the direction dependent Faraday component in $\jones_{pix}$, and thus  the normalisation matrix, but this would involve fitting a time and direction dependent Faraday rotation, which we have not attempted here.
Another consideration is that $\jones_{pix}$ is not calculated for every timestep, instead it is assumed to be valid for a length of time, typically  a few minutes. Simulations indicate that the validity of this approximation is a strong function of hour angle, with observations taken at zenith, immune to this effect for all reasonable integration times, but observations at 30 degrees zenith angle will suffer 1 \% of polarization leakage for integrations of 2 minutes.  To counter this effect in the 512T system the $\jones_{pix}$ will be updated every cadence (8s).

\section{The Observations}

\subsection{The Real-Time System Constraints}

As discussed in the introduction, the MWA when fully deployed will  run in real-time and therefore require considerable computing resources. The MWA real-time computer (RTC), as implemented in the MWA-32T prototype presages the architecture of the planned full RTC. It consists of a cluster of four, quad-core Xeon E54 servers, each containing an nVidia GeForce 8800 GTS graphics card (GPU). Visibilities from the correlator are presented to the cluster via a gigabit ethernet network. Each server receives 192 consecutive 40~kHz channels. These are processed individually and the entire imaging pipeline is run on the GPU. The processing load can be reduced by combining gridded visibilities across consecutive channels before imaging. Although the processing load on the calibration is a function of the number of baselines, the imaging load is a function of the number of pixels, therefore the imaging pipeline is required to do a similar amount of work in the prototype as the full MWA. The full MWA will be higher resolution, resulting in more pixels, but in actuality the imaging load on an individual processing node in the 32T is higher than the full MWA because 192 channels are sent to each node instead of 12; in order to perform in real time it was necessary to make some adjustments to the pipeline. Firstly each imaging channel is $\sim$2MHz wide, this is much wider than the 40kHz required for 32T and was chosen to be the widest that was possible without suffering more than 1\% bandwidth de-correlation and is integrated for just over 3 minutes in memory before each snapshot is output to disk. 

\begin{figure}[htbp] 
   \centering
   \plotone{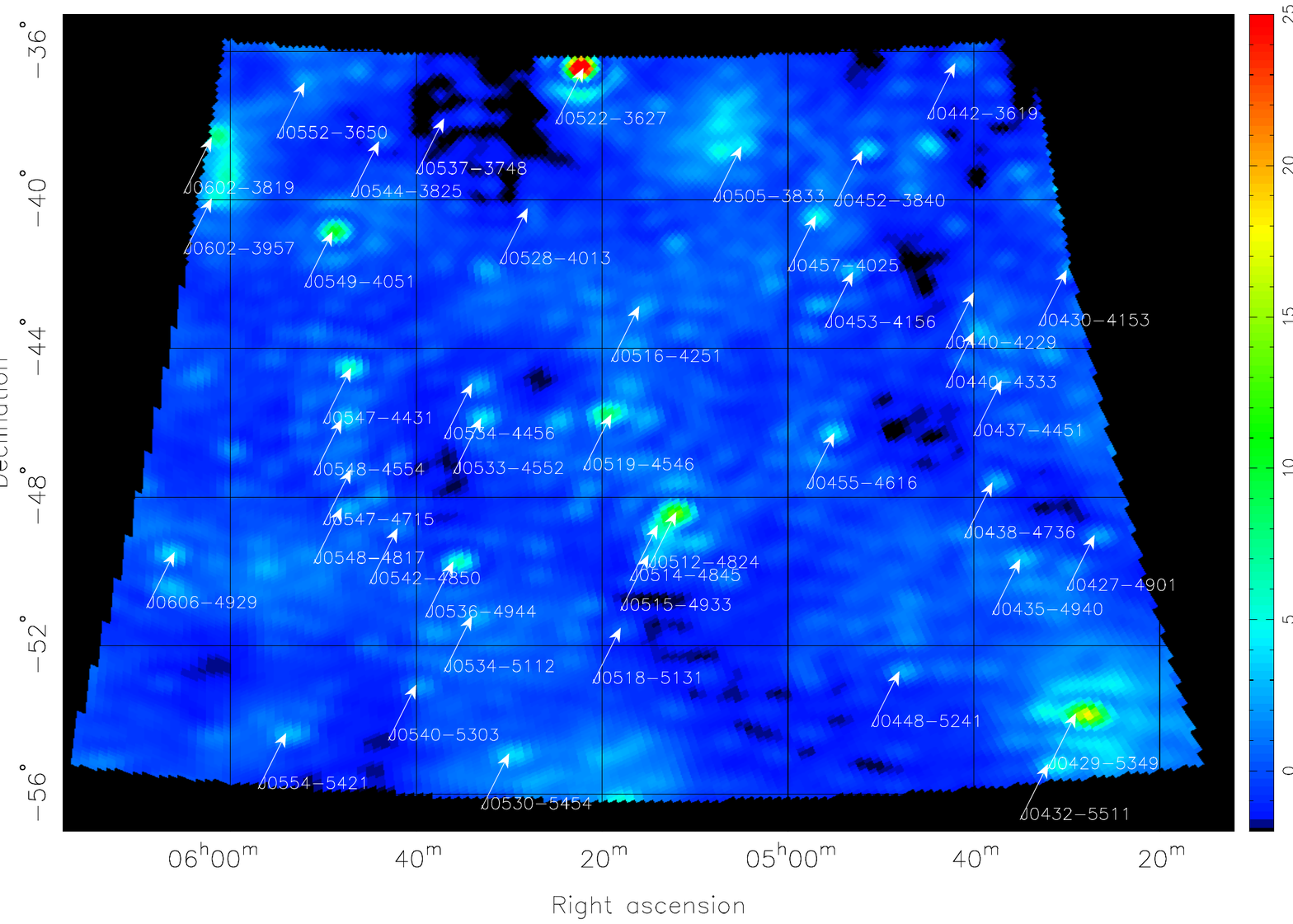} 
   \caption{Total intensity (Stokes I) image in Jy/beam with a central frequency of approximately 103 MHz, a bandwidth of 30 MHz and an integration time of approximately 4 hours. The pixelisation is HEALPIX, but the pixels have been rendered into a plate carr$\mathrm{\acute{e}}$e projection for display purposes. A selection of sources from the Parkes catalogue (\cite{pks90}) predicted to be above 5\,Jy in this field are labelled. Many more sources are visible than predicted the predominant reason for which is that they are predicted to have fluxes less than the 5\,Jy limit for the labelling.  The presence of Pictoris A (J0519-4546) residual in this image, despite an attempt to remove this 400\,Jy source from the visibilities, is indicative of an imperfect calibration. The grey scale has been clipped at 25\,Jy/beam to enhance contrast. The true peak pixel flux is approximately 60 Jy/beam and is associated with the point source in the North of the map, J0522-3627.}
   
 \label{fig:StokesI_low}
\end{figure}

\begin{figure}[htbp] 
   \centering
   \plotone{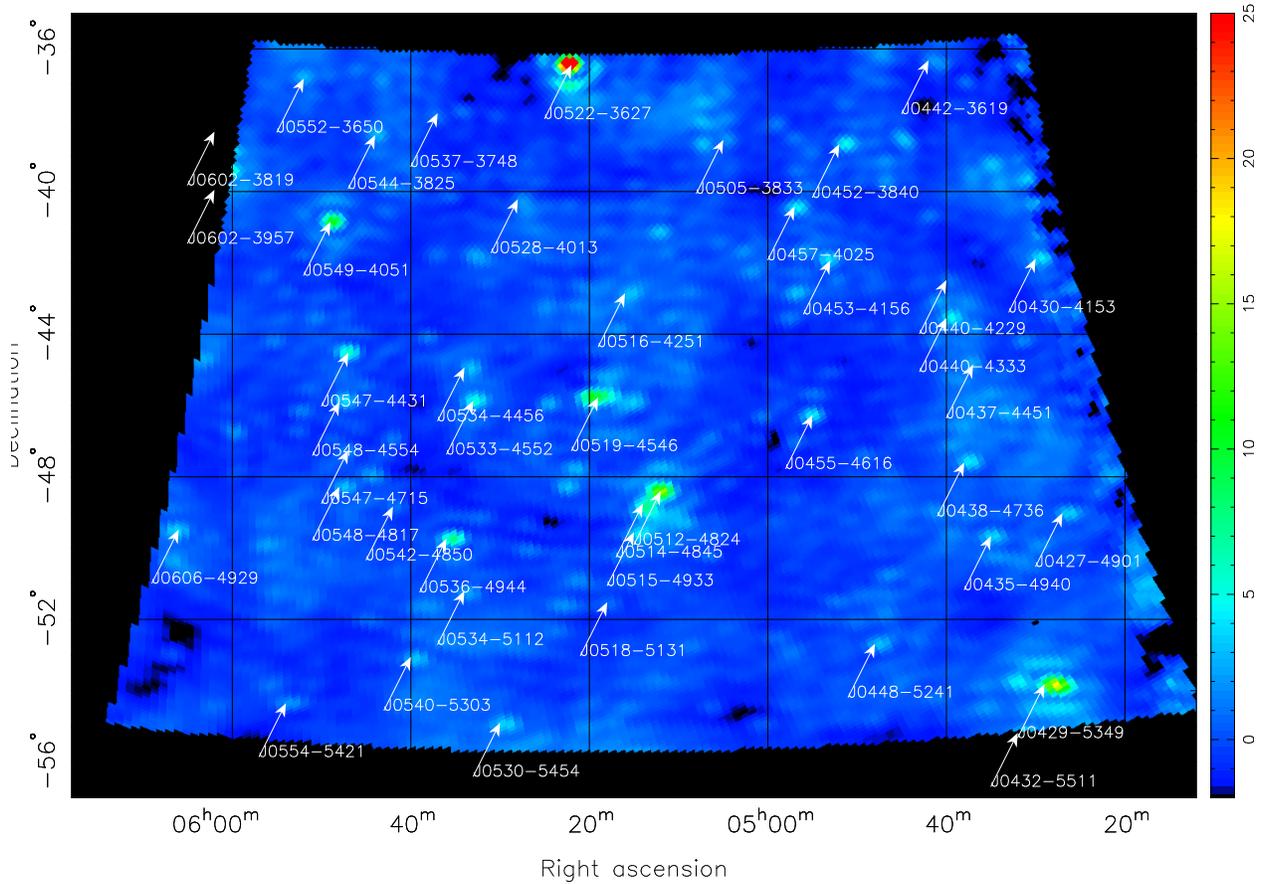} 
   \caption{Total intensity (Stokes I) image in Jy/beam with a central frequency of approximately 134 MHz, a bandwidth of 30 MHz and an integration time of approximately 4 hours. The pixelisation is HEALPIX, but the pixels have been rendered into a plate carr$\mathrm{\acute{e}}$e projection for display purposes. Sources from the Parkes catalogue predicted to be above 5Jy in this field are shown.  The presence of a Pictoris\,A (J0519-4546) residual in this image, despite an attempt to remove this 400\,Jy source from the visibilities, is indicative of an imperfect calibration. The grey scale has been clipped at 25\,Jy/beam to enhance contrast. The true peak pixel flux is approximately 60 Jy/beam and is associated with the point source in the North of the map, J0522-3627. }
   
 \label{fig:StokesI_mid}
\end{figure}

\begin{figure}[htbp] 
   \centering
   \plotone{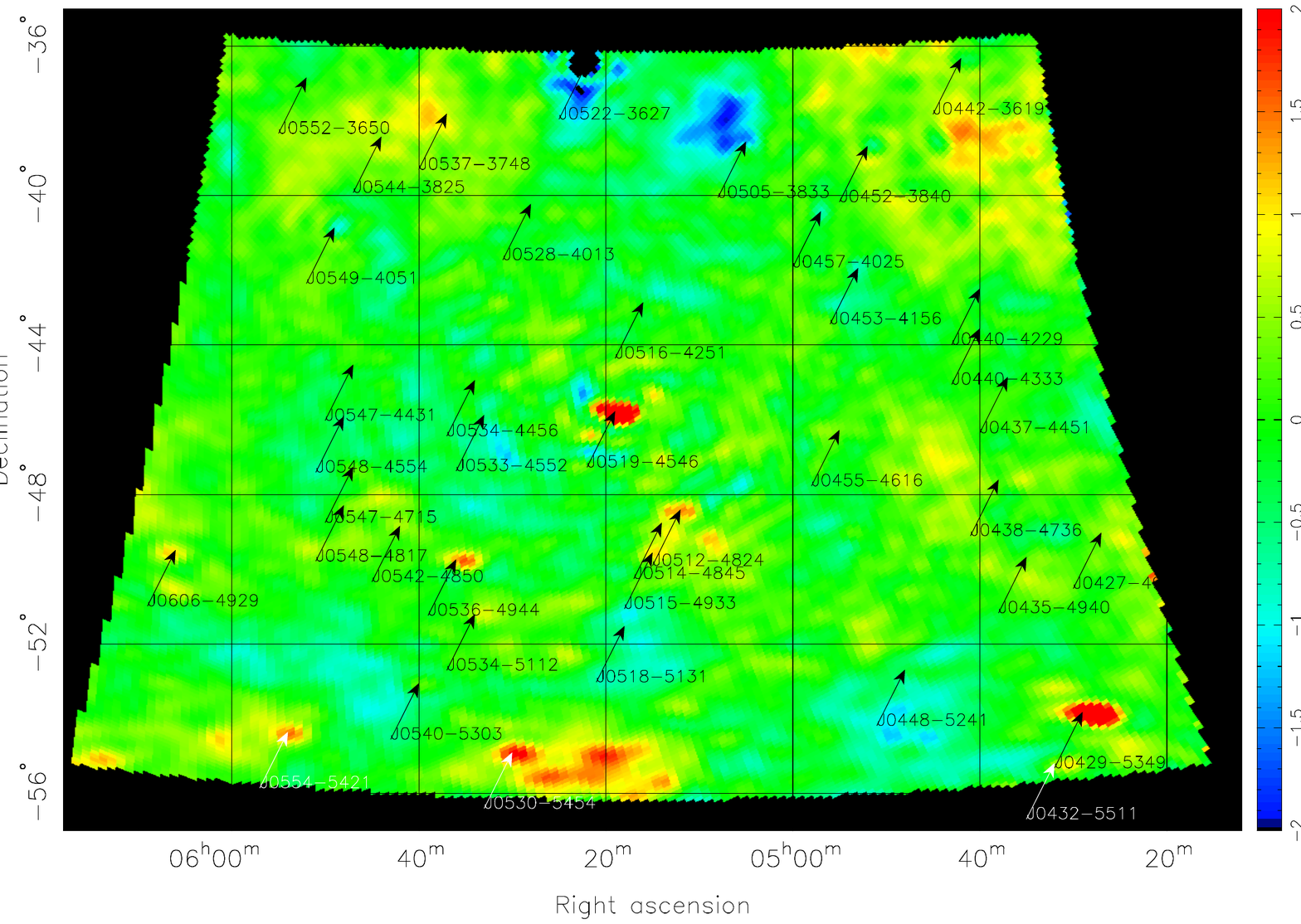} 
   \caption{The Stokes Q  image associated with the 103 MHz Stokes I image in Figure \ref{fig:StokesI_low}. The average brightness of the image is significantly lower than the Stokes I (0.14\,Jy/beam versus 0.9\,Jy/beam;  see Table \ref{tab:brightness}), but clear structure is evident associated with areas of high Stokes I. This is also evidence of calibration errors at the level of a few percent of Stokes I. As discussed in the text the low frequency and wide bandwidth make it unlikely that the polarization structure is intrinsic to the objects.  }
   \label{fig:StokesQ_low}
\end{figure}

\begin{figure}[htbp] 
   \centering
   \plotone{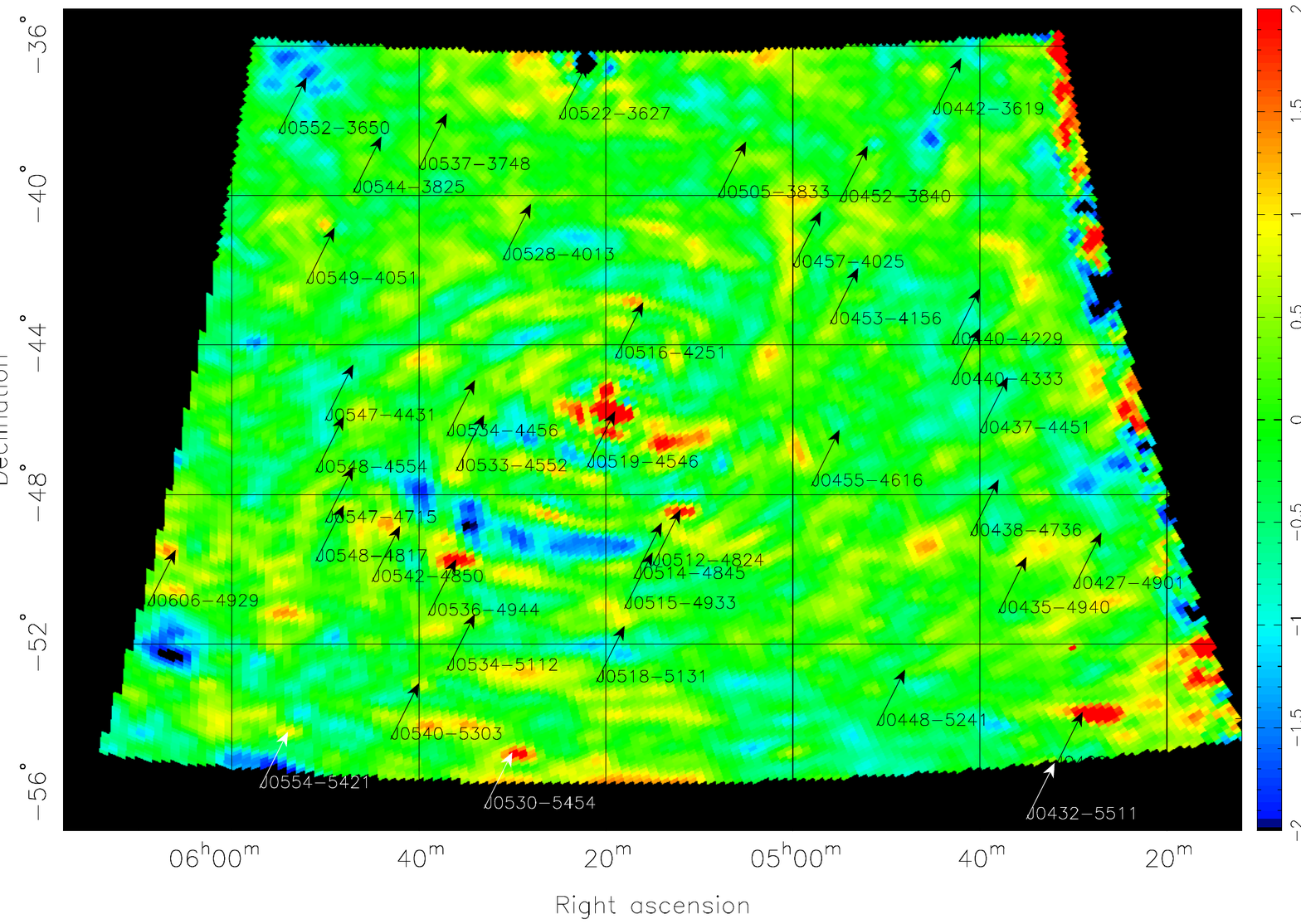} 
   \caption{The Stokes Q image associated with the 134 MHz Stokes I image in Figure \ref{fig:StokesI_mid}. The average brightness of the image is significantly lower than the Stokes I ( 0.18\,Jy/beam versus 0.76\,Jy/beam; see Table \ref{tab:brightness}) but clear structure is evident associated with areas of high Stokes I. This is also evidence of calibration errors at the level of a few percent of Stokes I. As discussed in the text the low frequency and wide bandwidth make it unlikely that the polarization structure is intrinsic to the objects.  }
   \label{fig:StokesQ_mid}
\end{figure}

\begin{figure}[htbp] 
   \centering
   \plotone{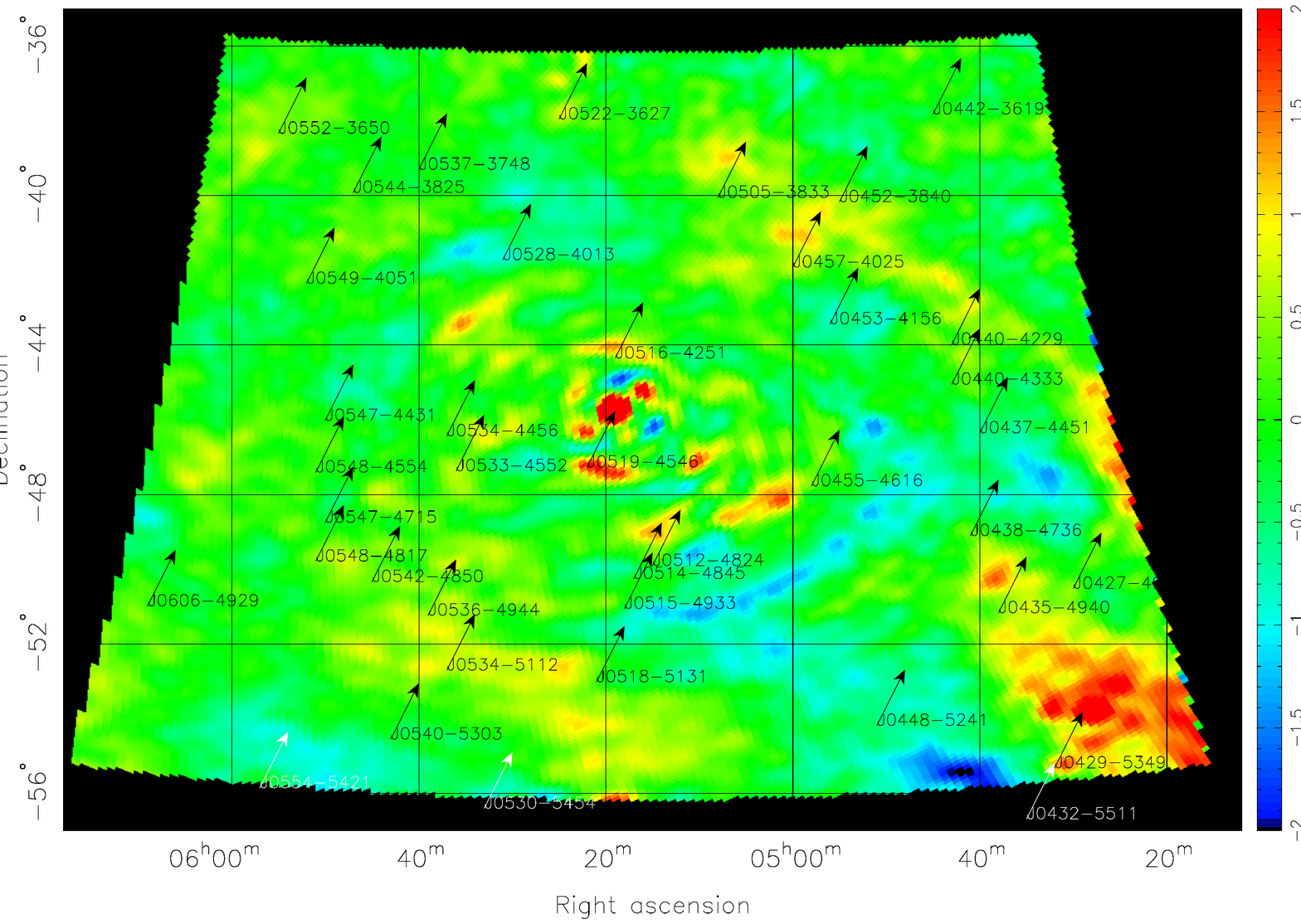} 
   \caption{The Stokes U image associated with the 103 MHz Stokes I image in Figure \ref{fig:StokesI_low}. The root mean square brightness of the image is significantly lower than the Stokes I (0.16\,Jy/beam versus 0.9\,Jy/beam; see Table \ref{tab:brightness}). There is considerably less association between bright Stokes I points and features in the Stokes U map, than evident in Stokes Q, although J0429-5349 is clearly present.}
   \label{fig:StokesU_low}
\end{figure}

\begin{figure}[htbp] 
   \centering
   \plotone{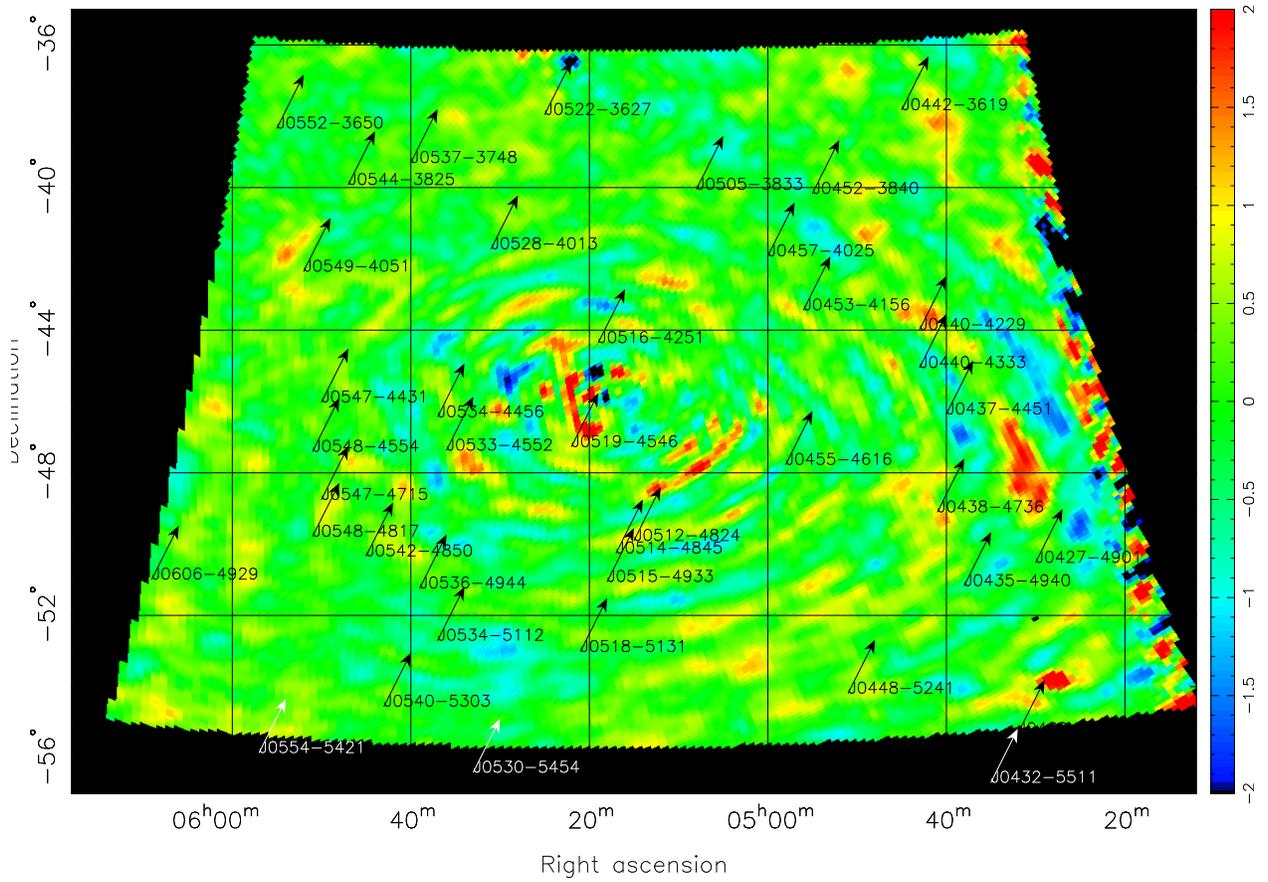} 
   \caption{The Stokes U image associated with the 134 MHz Stokes I image in Figure \ref{fig:StokesI_mid}. The root mean square brightness of the image is significantly lower than the Stokes I (0.19\,Jy/beam versus 0.76\,Jy/beam; see Table \ref{tab:brightness}). There is considerably less association between bright Stokes I points and features in the Stokes U map, than evident in the  Stokes Q map although J0429-5349 is clearly present.}
   \label{fig:StokesU_mid}
\end{figure}

As part of an MWA engineering test program in January 2010, multiple tracks of a 20$^\circ$ field around Pictor A were calibrated and imaged by the MWA real-time system. We present two 20$^\circ$ wide observations of Stokes I, Q and U centered on Pictor A, across ~30\,MHz and over $\sim$4 hours; which were taken on the 18th and 19th of  January 2010. The RTS generated a calibration data stream and wide-field sky images with an 8 second cadence;  calibration, imaging  and integration to 3 minutes was done in real-time. The RTS output was written in the specific projection and all-sky storage frame that has been selected for the MWA: HEALPIX, and has been weighted for optimal averaging in time and frequency. The output images have been subsequently integrated off-line, primary beam corrected and converted to Stokes parameters to produce the images presented in Figures \ref{fig:StokesI_low} to \ref{fig:StokesU_mid}.

\subsection{Dynamic Range and Peeling}

The noise in the integrated maps presented here, both simulated and actual, is dominated by sidelobe contamination and not thermal noise. Pictoris\,A is approximately 400Jy at these frequencies and its sidelobes are the main contributor to noise in unprocessed images. We have measured the root--mean--square brightness in Jy/beam for each of the images and present them in Table \ref{tab:brightness}. As discussed in \S \ref{sec:cal}, the 32T version of the RTS is based around 
measuring a Jones matrix for each tile towards a primary calibrator. 
This serves two purposes: since it is assumed that the tiles have known 
and identical beam shapes, the measured Jones matrices can be used with 
a source model to calibrate the visibilities over the entire primary 
beam; and a calibrated model of the source and its sidelobes can be 
subtracted (or {\em peeled}) from the visibilities before gridding and imaging. We have subtracted a point source model of Pictor A directly from the visibilities during the calibration process in real-time.  The 
low signal-to-noise ratio (SNR) for a set of instantaneous (8\,s) 32T 
visibilities limits the accuracy of our calibrator model, and as a 
result a clear residual  can be 
seen after the image noise has been integrated down. However this process has increased the dynamic range displayed in these images by a factor of 10, to about 400, the subtraction of adjacent channels, which suppresses the sidelobe noise very effectively, results in a noise floor consistent with a dynamic range of 1800.

Iterative deconvolution as it is routinely employed in interferometric imaging is not possible within this pipeline. The point spread function of the instrument is spatially variant, and there is no access to the visibilities. It is for this reason that we have employed peeling  to remove point sources from the ungridded visibilities. The 32T prototype lacks the sensitivity to remove many sources this way, and the full array will be limited not by sensitivity but by how many sources can be peeled in the allotted time. As a result there will be residual point sources and polarized diffuse emission that will need to be removed by post-processing and it is likely that sky modeling plus simulations of the instrument response will be required to achieve this (G. Bernardi submitted, B. Pindor submitted).

\begin{table}[htbp]
   \centering
   \begin{tabular}{@{} lcr @{}} 
 
      Stokes Parameter  & Frequency (MHz) & RMS (Jy/beam)\\
      \hline
      I      & 103 & 0.9 \\
      I	    & 134 & 0.76 \\
      Q	    & 103 & 0.14 \\
      Q    & 134  & 0.18 \\
      U	    & 103 & 0.16 \\
      U    & 134 &  0.19 
   \end{tabular}
   \caption{The root--mean--square brightness in Jy/beam of the observations presented. The integration time is approximately 4 hours and bandwidth approximately 30 MHz, there is some variation between integrations due a variable amount of frequency and time data flagging, which fractionally lowers the effective bandwidth and integration time.}
   \label{tab:brightness}
\end{table}


\subsection{Comparison to Simulations}

\begin{figure}[htbp] 
   \centering
   \plotone{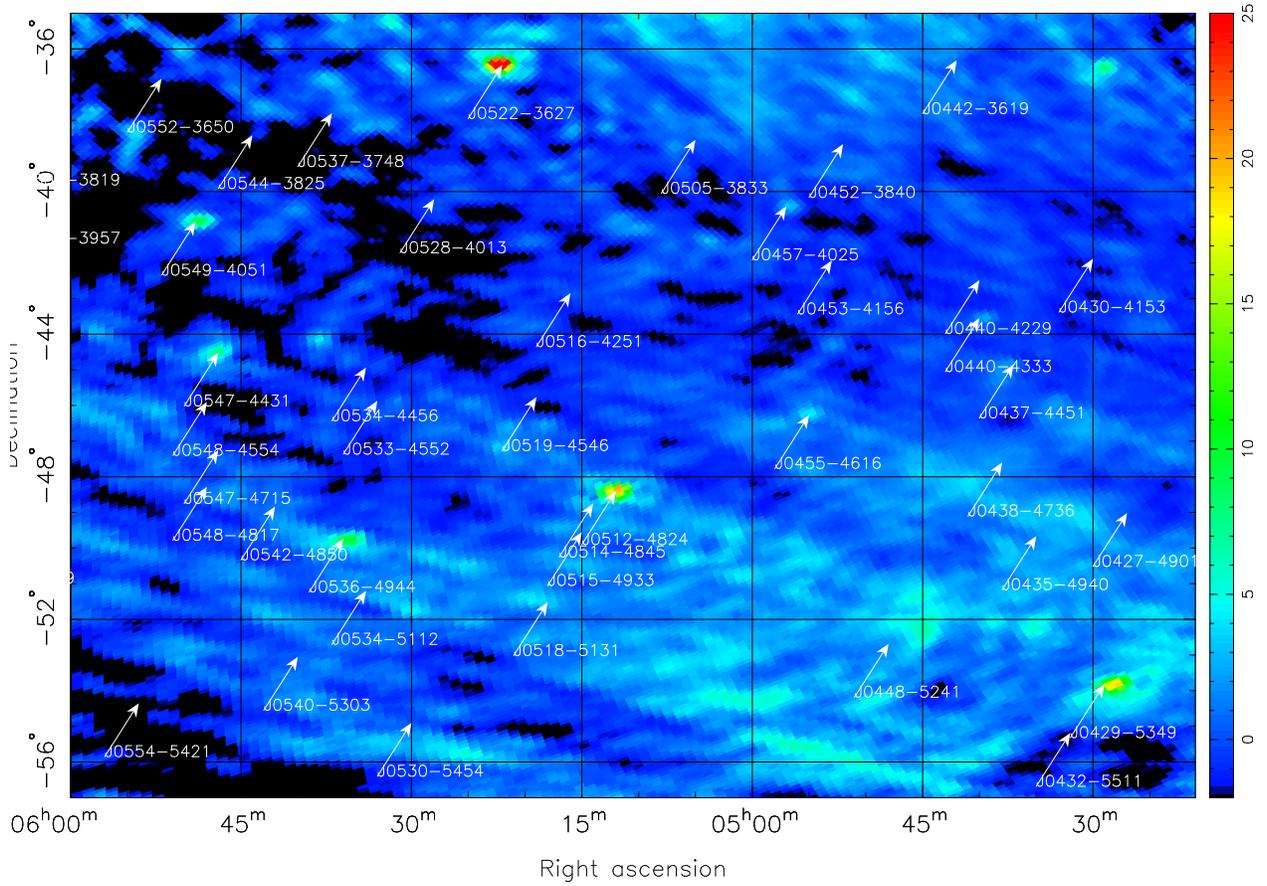} 
   \caption{Total intensity (Stokes I) image of a simulated sky at a central frequency of approximately 160 MHz, with a bandwidth of 40 kHz and an integration time of approximately 2 hours. The pixelisation is HEALPIX, but the pixels have been rendered into a plate carr$\mathrm{\acute{e}}$e projection for display purposes. The color scale has been clipped at 25Jy/beam to enhance contrast. The true peak pixel flux is approximately 60 Jy/beam and is associated with the point source in the North of the map, J0522-3627.}
   
 \label{fig:mapsI}
\end{figure}

\begin{figure}[htbp] 
   \centering
   \plotone{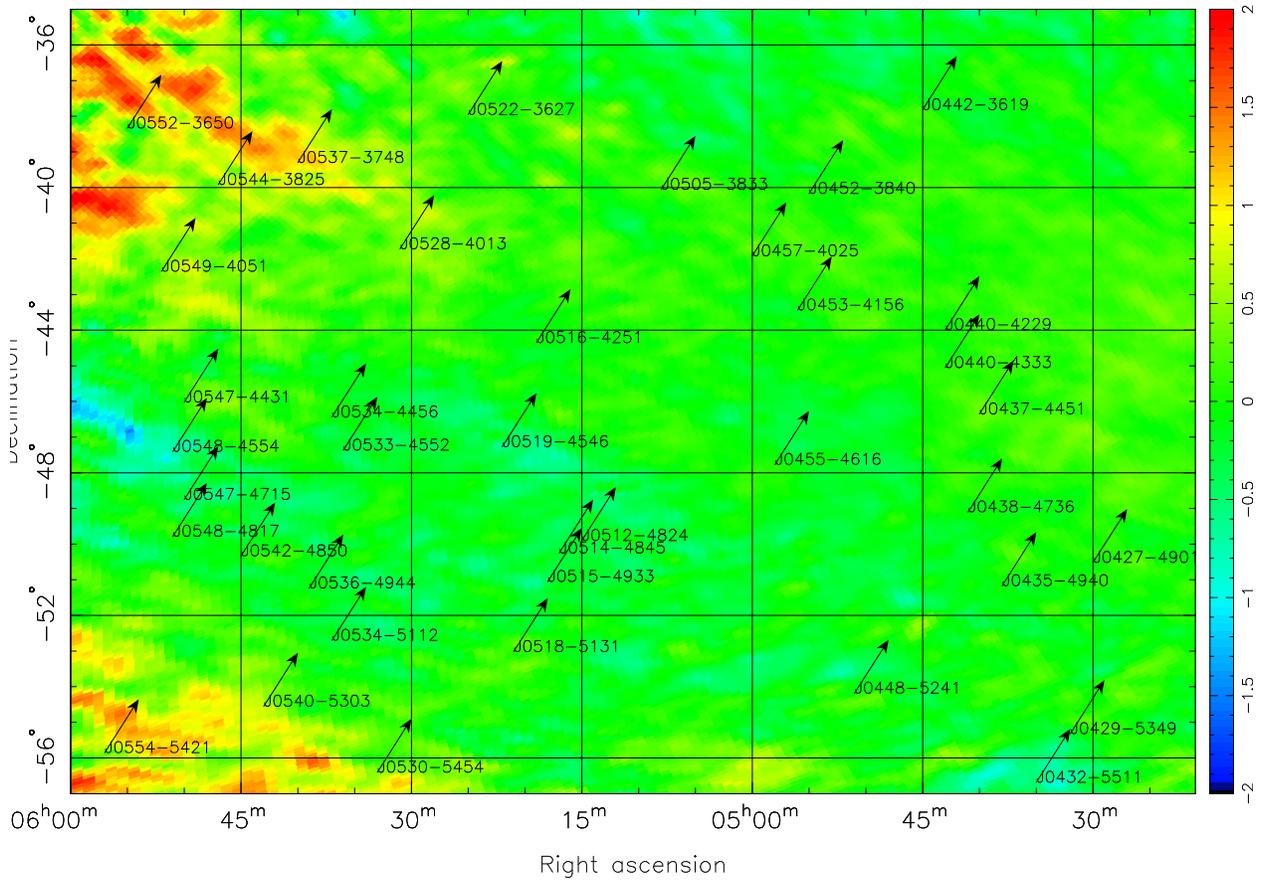} 
   \caption{The Stokes Q image of a simulated sky at approximately 160 \,MHz, with a bandwidth of 40\,kHz and an integration time of approximately 2 hours. The root mean square brightness of the image is significantly lower than the Stokes I image shown in Figure \ref{fig:mapsI},  this image shows considerably less association with high Stokes I features, but does display considerable diffuse polarization, this is due to the sidelobes of bright diffuse emission containing significant instrumental polarization.}
   \label{fig:mapsQ}
\end{figure}

\begin{figure}[htbp] 
   \centering
   \plotone{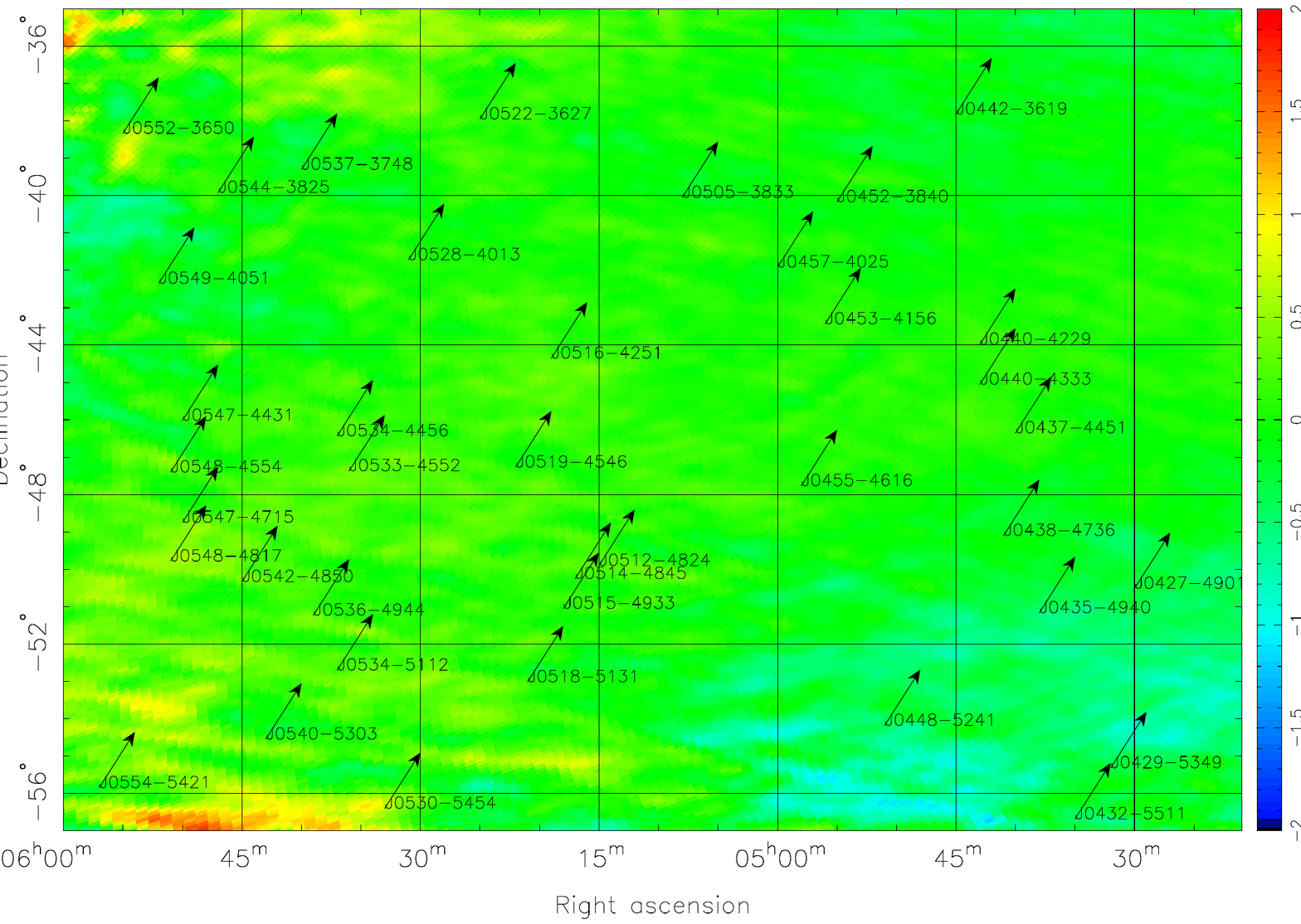} 
   \caption{The Stokes U image of a simulated sky, at a central frequency of approximately 160 MHz, with a bandwidth of 40 kHz and an integration time of approximately 2 hours. The root mean square brightness of the image is significantly lower than the Stokes I image, shown in Figure  \ref{fig:mapsI},  this image shows considerably less association with high Stokes I features, but does display considerable diffuse polarization, this is due to the sidelobes of bright diffuse emission containing significant instrumental polarization.}
   \label{fig:mapsU}
\end{figure}

The MAPS software package (R. B. Wayth et al. in preparation) has been  used to provide a well understood simulated input dataset to the software pipeline. Figures \ref{fig:mapsI}, \ref{fig:mapsQ} and \ref{fig:mapsU} are the companions to the actual observations presented, although at a slightly higher frequency (160.02\,MHz). The simulation contains a subset of the Parkes catalogue of point sources, and a diffuse background in Stokes I from \cite{hks+81}. The simulator generates visibilities as they would be produced by the MWA correlator, and the array parameters can be controlled by the user. In this case the calibration is perfect, there is no ionospheric refraction, and all the point sources and diffuse background are intrinsically unpolarized. The integration time was approximately 2 hours and only a single 40\,kHz channel centered at 160.02 MHz was simulated. The full observation has not been simulated as the computational demands are large and a representative integration is required to test the fidelity of the Stokes conversion. The two hour integration is sufficient to ensure that the dominant noise source in the simulated map is from sidelobes. The images were produced by exactly the same pipeline as produced the real sky images, except that these images were generated offline and not in real time. The FOV imaged was also slightly wider: closer to 30$^\circ$. Pictoris\,A has been peeled out the simulated images, due to  perfect calibration there are no residuals remaining.

Figures \ref{fig:StokesI_low} and  \ref{fig:StokesI_mid}  contain significantly more point sources than the simulation (Figure \ref{fig:mapsI}).  These sources are actually in the catalog. However the simulation is very conservative when including catalog sources, and has only included those sources with flux measurements near the MWA observing band and that are predicted to be larger than 5\,Jy. Nevertheless this simulation gives a good indication of the performance of the system under more controlled conditions than the measured dataset. 

\subsubsection{Polarization Properties}

The RMS brightness of the polarization observations is approximately 16\% of the Stokes I observations, the point source polarized residual is considerably less than this, 9\% of J0522--3627 at the very edge of the Q map, and only 2\% at the same point in the U map. The residual is also a function of position within the primary beam, near the center of the map the residual in Q is nearer 2\% and in the U maps there is almost no measurable association between Stokes I sources and polarized structure in the beam center - except as polarized residuals from the peeling of Pictoris\,A. The polarized residuals are a feature of imperfect calibration of the individual antennas. The fact that the residuals deteriorate away from beam center lends further weight to this claim; as differences between the primary beams become more apparent further from the beam center.

The simulated point sources have no intrinsic polarization and as the calibration is perfect there should be no polarized point sources in the integrated maps, unlike the images of the true sky which do display some polarized point source signals due to imperfect calibration of the primary beams. Figures {\ref{fig:mapsQ} and \ref{fig:mapsU} are featureless in this regard, with little evidence of the existence of any point source features in the maps. The maps from simulation possess diffuse polarization despite the fact that the input sky is unpolarized.  This is due to the polarized sidelobes of the Stokes I emission. The polarized brightness in a given pixel contains contributions from not only the polarized sky at that position, but all the sidelobes of the diffuse and point source emission in the sky. The sidelobes are generated with instrumental polarization commensurate with the position of the generating source, not the position at which the sidelobe is measured. This response is modulated by the antenna gain so sources outside the primary beam will be suppressed,  however bright point sources, and patches of diffuse emission, will still display polarized sidelobes across the images, these will generate apparent polarized signal where none exists. This produces an error in the Stokes conversion even if the array is perfectly calibrated. This is the prevailing polarized signal in images of the simulated sky and is no doubt also present in the images of the true sky, but the significant frequency synthesis over the 30MHz bandwidth mitigates the effect and therefore calibration errors dominate those images.

This effect will severely limit the polarization fidelity of integrated images unless an accurate model of the radio sky can be obtained. A significant project is underway with the MWA to obtain an initial model of the sky through an all-sky survey project that will be continually improved through the lifetime of the instrument.

\section{Conclusions}

We have demonstrated the performance of a real-time calibration and imaging pipeline for the MWA-32T. Wide--field polarimetric imaging has successfully been performed via the integration of warped, and weighted snapshots. The simulations indicate that the calibration and imaging pipeline works with sufficient fidelity to remove instrumental polarization from a simulated point source population. But the observed point source population displays residual polarization, in the worst case, at the $<$ 10\% level and in general at the $<$ 5\% level; this is likely due to an imperfect calibration. Calibration precision is limited by low snapshot sensitivity and the fact that we cannot account for the differing primary beam shapes of array elements.  The simulations also indicate that a major component of measured polarized emission from this and other wide-field-of-view instruments will be due to the polarized sidelobes of bright unpolarized features observed off-axis; necessitating the construction of comprehensive sky models applicable to these instruments. The removal of which, preferably in the visibility domain, or via an iterative forward modeling scheme being required to limit this instrumental polarization.

The MWA elements are crossed dipoles on a ground-plane, electrically phased to a pointing direction and have considerable instrumental polarization (up to ~100 \%). We have removed instrumental polarization down to $<$ 5 \% in real time; with limited sensitivity and a common scaled model for each array element response. The 512 tile MWA, with better constrained, individual tile responses and much greater snapshot sensitivity should perform substantially better. The key improvement lies not simply in improving sensitivity but accounting for the different, direction dependent, polarimetric responses of the constituent antennas.

Subsequent publications in this series will aim to perform point source subtraction from these images, and others like them at different frequencies, in order to obtain spectral index information; and to constrain the behaviour of the MWA antenna beam.

\section*{Acknowledgments} 

This work uses data obtained from the Murchison Radio-astronomy Observatory. We acknowledge the Wajarri Yamatji people as the traditional owners of the Observatory site. Support came from the U.S. National Science Foundation (grants AST-0457585 and PHY-0835713), the Australian Research Council (grants LE0775621 and LE0882938), the U.S. Air Force Office of Scientific Research (grant FA9550-0510247), the Smithsonian Astrophysical Observatory, the MIT School of Science, the Raman Research Institute, the Australian National University, the iVEC Petabyte Data Store, the Initiative in Innovative Computing and NVIDIA sponsored Center for Excellence at Harvard, and the International Centre for Radio Astronomy Research, a Joint Venture of Curtin University of Technology and The University of Western Australia, funded by the Western Australian State government.

\bibliography{pipeline}

\bibliographystyle{apj}

\end{document}